\documentclass{article}

\usepackage{arxiv}

\usepackage{graphicx}

\usepackage[utf8]{inputenc} 
\usepackage[T1]{fontenc}    
\usepackage{hyperref}       
\usepackage{url}            
\usepackage{booktabs}       
\usepackage{amsfonts}       
\usepackage{nicefrac}       
\usepackage{microtype}      
\usepackage{lipsum}

\title{A Multimodal CNN-based  Tool  to Censure Inappropriate Video Scenes}

\author{
 Pedro V. A. de Freitas \\
 PUC-Rio\\
 \texttt{pedropva@telemidia.puc-rio.br}
 \And
 Paulo R. C. Mendes \\
 PUC-Rio\\
 \texttt{paulo.mendes@telemidia.puc-rio.br}
 \And
 Gabriel N. P. dos Santos \\
 PUC-Rio\\
 \texttt{gabrielpereira@telemidia.puc-rio.br}
 \And
 Antonio Jos\'e G. Busson \\
 PUC-Rio\\
 \texttt{busson@telemidia.puc-rio.br}
 \And
 \'Alan Livio Guedes \\
 PUC-Rio\\
 \texttt{alan@telemidia.puc-rio.br}
 \And
 S\'ergio Colcher \\
 PUC-Rio\\
 \texttt{colcher@inf.puc-rio.br}
 \And
 Ruy Luiz Milidi\'u \\
 PUC-Rio\\
 \texttt{milidiu@inf.puc-rio.br}
 }

\begin{document}
\maketitle

\begin{abstract}
Due to the extensive use of video-sharing platforms and services for their storage, the amount of such media on the internet has become massive. 
This volume of data makes it difficult to control the kind of content that may be present in such video files. 
One of the main concerns regarding the video content is if it has an inappropriate subject matter, such as nudity, violence, or other potentially disturbing content. 
More than telling if a video is either appropriate or inappropriate, it is also important to identify which parts of it contain such content, for preserving parts that would be discarded in a simple broad analysis. 
In this work, we present a multimodal~(using audio and image features) architecture based on Convolutional Neural Networks (CNNs) for detecting inappropriate scenes in video files. 
In the task of classifying video files, our model achieved 98.95\% and 98.94\% of F1-score for the appropriate and inappropriate classes, respectively. 
We also present a censoring tool that automatically censors inappropriate segments of a video file.

\end{abstract}


\section{Introduction}

The amount of multimedia content on the internet, especially video, is increasing each year.
More than 300 hours of video are uploaded to YouTube every minute\footnote{\url{https://biographon.com/youtube-stats/}}. 
Due to the amount of material uploaded, controlling the content that is loaded is quite challenging even for large companies.
For example, Facebook and Youtube are being sued for hosting videos from the Christchurch shootings\footnote{\url{https://www.bbc.com/news/technology-47705904}}.

\begin{figure}[ht] 
    \centering
\includegraphics[width=0.55\textwidth]{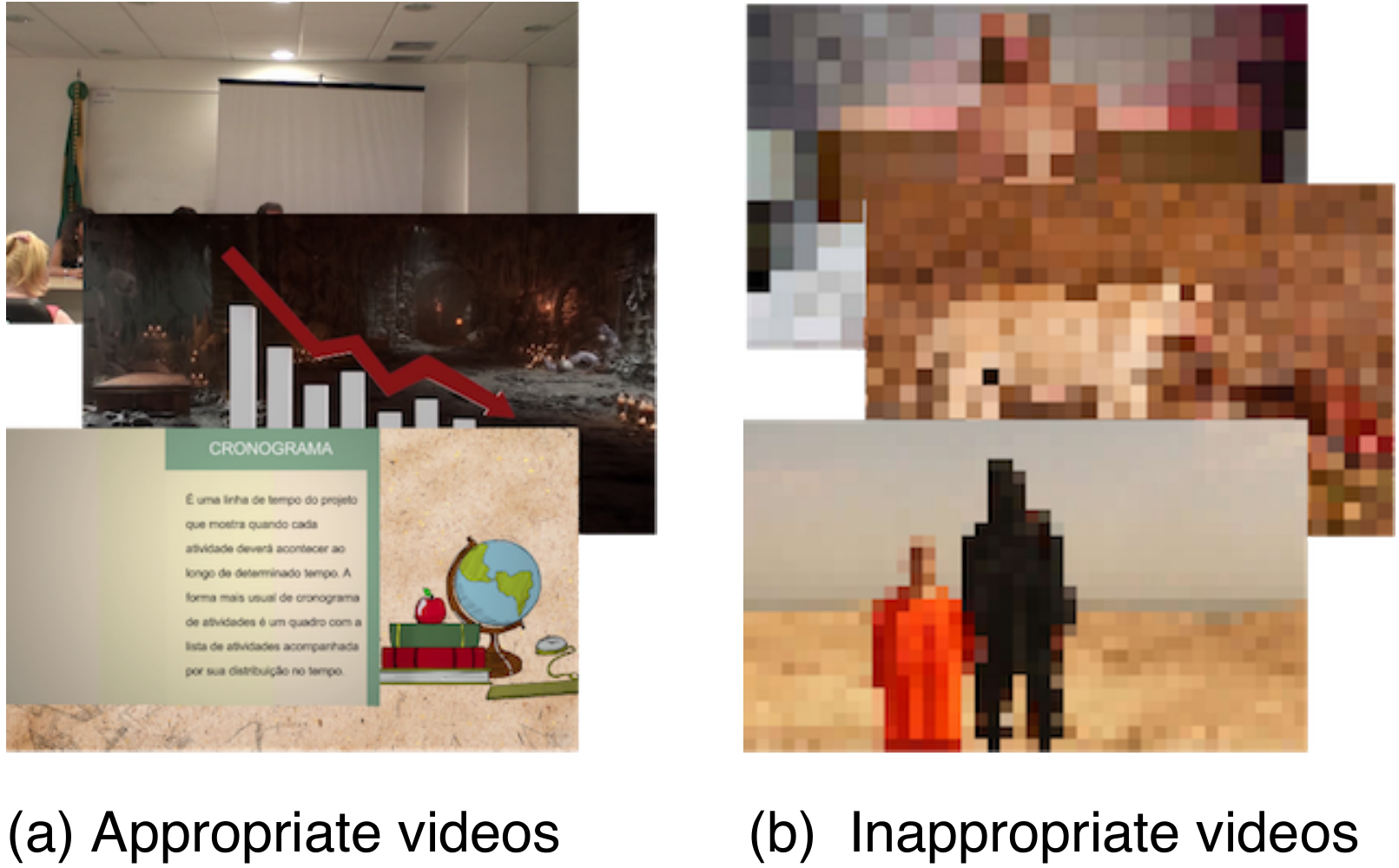}
    \caption{Examples of each category}
     \label{fig:samples}
\end{figure}

The word \emph{Inappropriate} is often used as a reference to media that contain content such as nudity, intense sexuality, violence, gore or other potentially disturbing subject matter. 
On the other hand, \emph{Appropriate} means that a content is suitable for most viewers. Figure \ref{fig:samples} illustrates these two categories.
There are three scenes with appropriate content on the left, and three scenes with inappropriate on the right.

Besides controlling the content of entire video files, it is also important to identify which scenes of it are inappropriate. 
For example, YouTube only considers the entire video for telling if it is either appropriate or inappropriate--only one inappropriate scene is necessary for banning an entire video from the platform. 
For that reason, an ongoing problem is that, in some cases, a video has its entire content labeled as inappropriate when only specific scenes contain inappropriate content (e.g. movies and documentaries on wars and conflicts). For instance, one can suppose that students are looking for a documentary about Vietnam War for their history class.
But they just cannot find any because the storage service in which they are searching has banned all documentaries in that subject for containing violent scenes.
Thus, a tool that not only labels videos but also provides information on which scenes are inappropriate would allow access to the video content, while preventing exposure to these scenes.  This work aims at presenting an approach for detecting and censoring inappropriate video scenes in video files.
This is done in a way that appropriate scenes are not censored and the inappropriate ones pass through a process to make them presentable. 

Controlling the type of content loaded to video storage service requires an automatic analysis efficiently and quickly.
Methods based on \textit{Deep Learning} (DL) became the \textit{state-of-the-art} in various segments related to automatic video analysis.
Convolutional Neural Networks (CNNs) architectures, or ConvNets, have become the primary method used for audio-visual pattern recognition.

Other works also share the motivation of classifying video files in the categories we mentioned~\cite{song2018pornographic,torres2018automatic,wehrmann2018adult}. 
However, most of them do not use audio and image for classification, or use hand-crafted feature extraction methods, or do not use the latest feature extraction CNNs, which have been showing great potential in video recognition and classification. 
Our work uses two deep CNNS, one to extract image sequence features and another to extract audio features. 
We combine those features to create a single feature vector for the entire video (or video segments), which then is used as input for the classifier. 
It is a rather simpler method for video classification and yet it still yields better results than the related work.

Similar to ours, \cite{tofa2017inappropriate} proposes a model for detecting inappropriate video scenes in video files. They extract the frames from the video file and use them as input to a CNN model to classify it. 
Their work differs from ours in the sense that they extract features only from images (video frames) and does not consider the audio track of the video. Instead, we divide the video into smaller video segments for extracting features from both its frames and audio track.

To present our proposal, this paper is organized as follows. 
Section \ref{sec:model} discusses the used model for classifying a video as either appropriate or inappropriate. 
Section \ref{sec:censorship} presents the tool we developed for censoring inappropriate video scenes and preserve the appropriate ones. 
Section \ref{sec:experiments} shows the results we obtained in both the video classification and with our censorship tool. 
Finally, Section \ref{sec:conclusion} presents our final remarks and future work.

\section{Classification Model}
\label{sec:model}

Our CNN-based classifier is composed of two modules. 
The first module is what researchers call the \textit{backbone}, which acts as a feature extractor from which the model draws its discriminating power.
The second module, the \textit{classifier}, operates over the extracted features by the backbone to aggregate and classify it. 
We opt for a bi-modal approach that uses two backbones to extract the audio and image features from videos. 
Once we have extracted the features from the video, we then use a shallow model to perform the video classification. 
In the remainder of this section, we detail the embeddings extractor and the algorithms used for classification.

\begin{figure*}[!t]
    \centering
    \includegraphics[width=\textwidth]{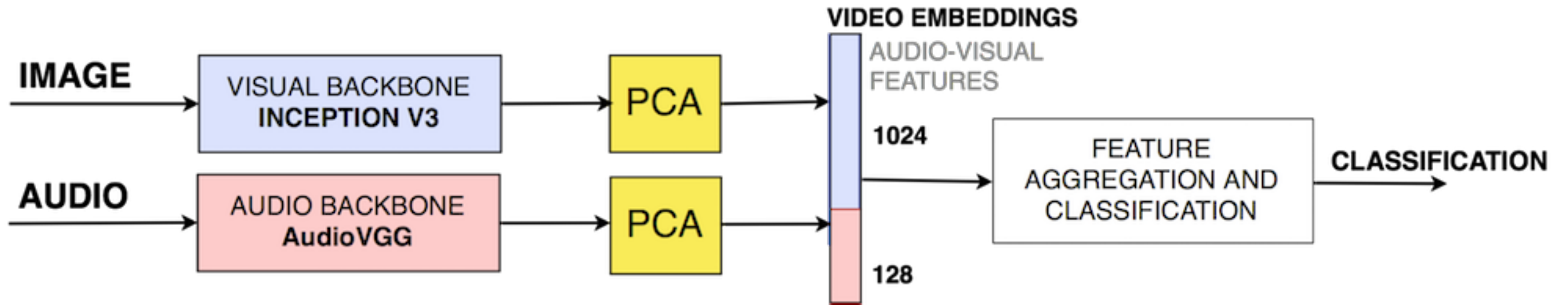}
    \caption{Multimodal architecture for inappropriate video classification}
    \label{fig:model}
\end{figure*}

A CNN, when trained, tends to learn at the first layers the low-level features (e.g. in the visual domain: edges, corners, contours). 
At the intermediate and final layers, the combination of these filters helps to extract more complexes features, resulting in a vector of continuous numbers called \textit{embeddings}. 
In this work, we use two benchmark CNNs to extract both image and audio \textit{embeddings} from videos by using the transfer learning technique~\cite{tan2018survey}.

First we extract both visual and audio \textit{embeddings}.
Based in the work of Abu-El-Haija \textit{et al.}~\cite{abu2016youtube}, we decode each video at 1 frame-per-second up to the first 360 seconds and feed an \textbf{InceptionV3}~\cite{szegedy2016rethinking} with the network weights pre-trained on ImageNet\footnote{\url{http://www.image-net.org/}} to extract the image \textit{embeddings}.
We also feed \textbf{AudioVGG}~\cite{hershey2017cnn} with the network weights pre-trained on Audioset\footnote{\url{https://research.google.com/audioset}} to extract the audio \textit{embeddings}. Next, we apply PCA (and whitening) to transform the dimensions of the image \textit{embeddings} to size 1024 and audio \textit{embeddings} to size 128. Finally, we concatenate both image and audio \textit{embeddings} to compose the final video \textit{embeddings} with 1152 dimensions.

The video \textit{embeddings} are then fed to a \textbf{Support Vector Machine (SVM)}~\cite{cortes1995support} for classification. 
In an SVM, the data is mapped into a higher dimension input space where an optimal separating hyper-plane is constructed. 
These decision surfaces are found by solving a linearly constrained quadratic programming problem. 
The architecture of our Inappropriate classifier is illustrated in Figure~\ref{fig:model}.

\subsection{Metrics}\label{sec:metrics}

We evaluate the model by the Precision (P), Recall (R) and F1-Score for appropriate and inappropriate classes:

    \begin{equation}
    \label{equation:precision}
    P = \frac{TP}{TP + FP}
    \end{equation}
    
    \begin{equation}
    \label{equation:recall}
    R = \frac{TP}{TP + FN}
    \end{equation}
    
    \begin{equation}
    \label{equation:f1}
    F1 = \frac{2 \times P \times R}{P + R}
    \end{equation}

Where $TP, TN, FP$ and $FN$ denote the examples that are true positives, true negatives, false positives, and false negatives, respectively. 
The F1 score, defined in Equation~\ref{equation:f1}, measures how precise the classifier by the harmonic mean between Precision (Equation~\ref{equation:precision}) and Recall (Equation~\ref{equation:recall}).
The F1-score represents an overall performance metric, and the precision and recall metrics can give insights on where the classification model is doing better.
\section{Censorship Tool}
\label{sec:censorship}

\begin{figure}[ht] 
    \centering
\includegraphics[width=0.5\linewidth]{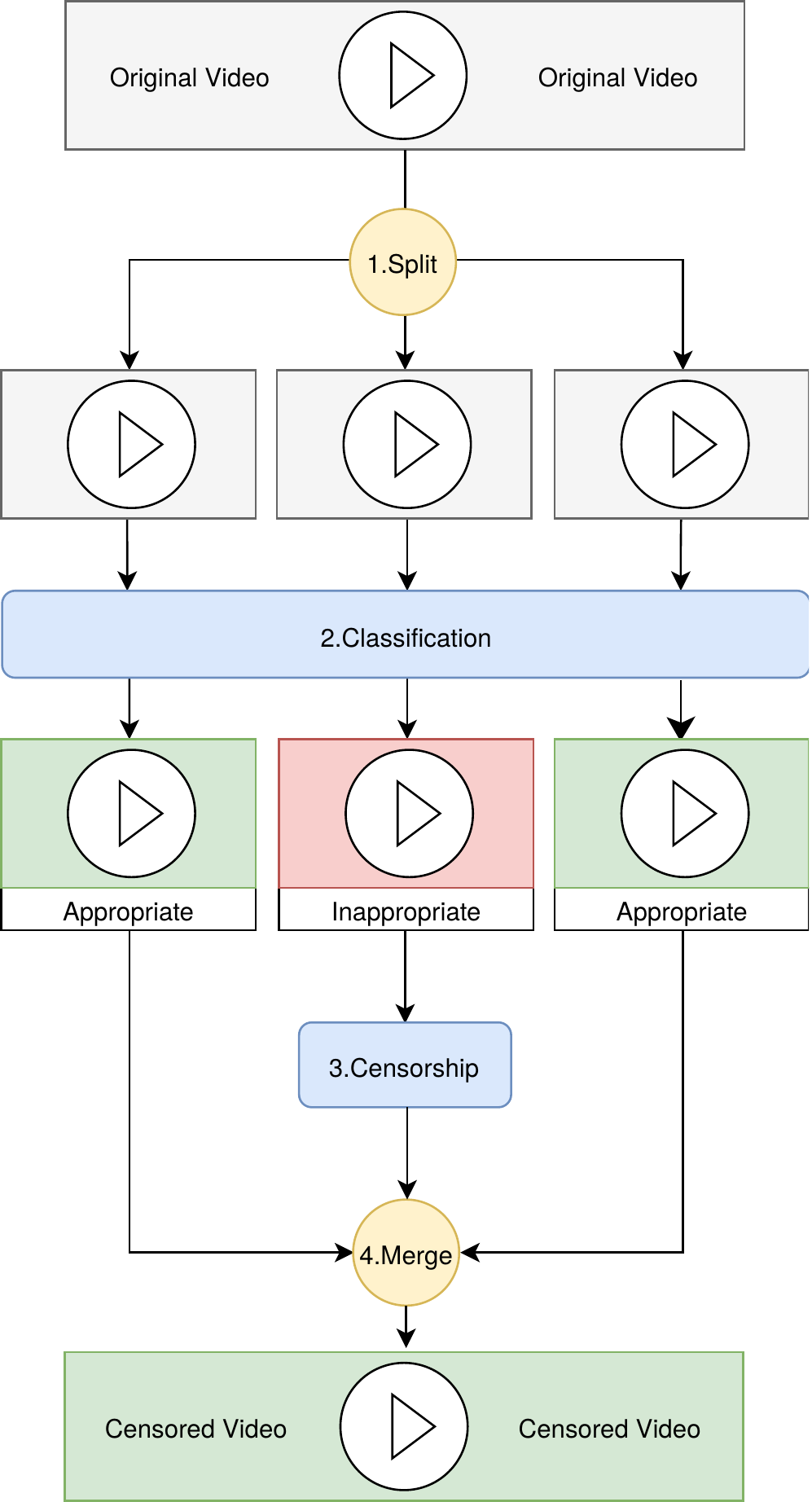}
    \caption{Censorship tool overview}
     \label{fig:imgtool}
\end{figure}

We designed a tool that receives a video and automatically censors parts of it that may contain inappropriate content. 
Figure \ref{fig:imgtool} shows how it works and such process is summarized in the following steps:

\begin{enumerate}
    \item \textbf{Split:} The video received is split into video segments (5 seconds each at most)
    \item \textbf{Classification:} Each segment is labeled in either appropriate or inappropriate.
    \item \textbf{Censorship:} If a segment is labeled as inappropriate, the audio of the video is removed and its image is blurred.
    \item \textbf{Merge:} Finally, the video segments are merged (the ones first labeled as appropriate and the ones that passed through the censorship process).
\end{enumerate}

Our censorship tool is implemented in the Python programming language. 
This choice comes from the fact that such language is broadly used for deep learning solutions and provides a myriad of libraries for audio, image and video processing. 

The \textit{Split} and \textit{Merge} steps are performed using the library MoviePy\footnote{\url{https://zulko.github.io/moviepy/}}. 
For the classification step, the classification model detailed in Section \ref{sec:model} is used. 

The \textit{Censorship} step is carried out by extracting all the frames of the video segment, applying a \emph{Gaussian Blur} for each frame, and then replacing the frames with the processed ones. 
The Gaussian Blur formula is the following:

\[G(x,y) = \frac{1}{2\pi \sigma ^2}e^{\frac{x^2+y^2}{2\sigma ^2}}\]

\begin{figure}[ht] 
    \centering
\includegraphics[width=0.4\linewidth]{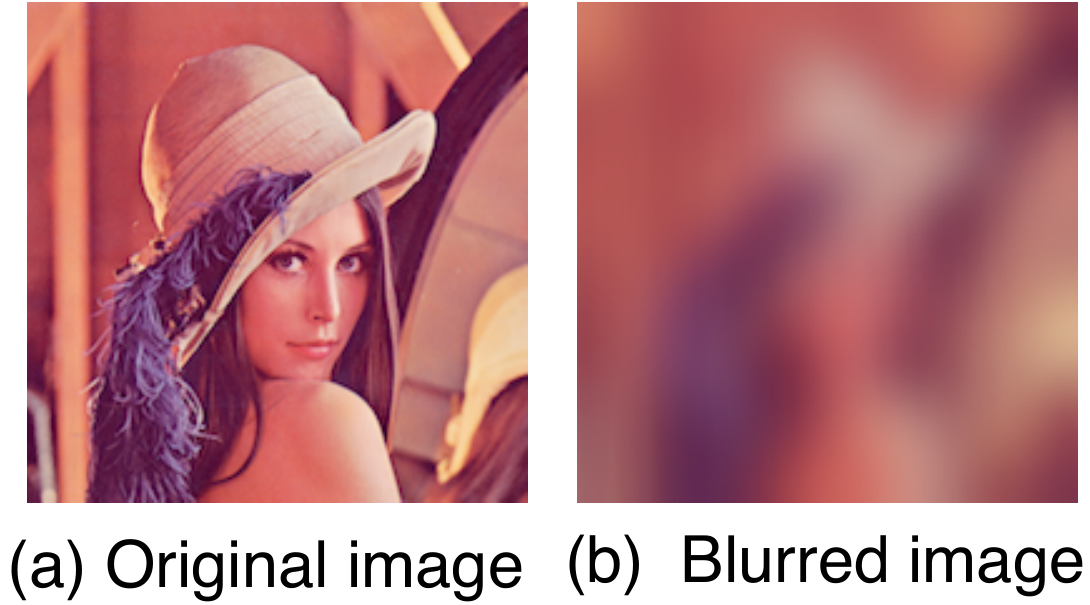}
    \caption{Example of Gaussian Blur}
     \label{fig:gaussian}
\end{figure}

In that formula, \(x\) is the distance from the origin in the horizontal axis, \(y\) is the distance from the origin in the vertical axis, and \(\sigma\) is the standard deviation of the Gaussian distribution. 
Figure \ref{fig:gaussian} shows an example of such image processing.

The audio track from the original video segment is not attached to the new one so that the audio censorship is performed by removing it.

Besides automatically censoring inappropriate video scenes from the video file, the tool we designed also returns an XML file that tells the beginning and duration of each inappropriate scene so that the video service has the autonomy to decide what to do with such scenes.

\section{Experiments}
\label{sec:experiments}

For model training, we used a video dataset with 33.500 inappropriate videos and 33.500 appropriate videos. 
The inappropriate video files consist of pornographic and violent content.

The pornographic content was obtained from the  XVideos\footnote{\url{https://info.xvideos.com/db}} database. 
This database was chosen as a source for pornographic content because of its size (7 million videos) and variety of annotations~(minor and major tags --each video has one major tag and many other minor tags). 
To select a sample of this database, we selected videos from the 70 major tags to maintain a similar distribution to the original database. 
In particular, to prevent lower-quantity major tags from disappearing, we have defined a minimum of 10 videos for each major tag.

The violent content (hereafter referred to as \textit{gore}) was collected from specialized websites through web crawlers (e.g. bestgore\footnote{\url{https://www.bestgore.com}}). 
It is mainly composed of videos depicting deaths, exposed injuries, diseases, accidents, and other mentally disturbing imaging.

As for the appropriate videos, we collected them from the Yotube8M\footnote{\url{https://research.google.com/youtube8m}} dataset. 
This choice comes from the size of the dataset (almost 8 million videos), the diverse tagging and the video classification challenges it holds, which make the dataset largely available.
We also added the Cholec80\cite{twinanda2016endonet} dataset to ours, it contains 80 videos of cholecystectomy surgeries performed by 13 surgeons. 
All videos from the Cholec80 dataset were labeled as appropriate since videos of surgery are usual in some specific contexts (e.g. educational context).


We split the dataset into 90\% for training and 10\% for test. 
Then, we perform a 20-fold cross validation and use \textit{f1-score}, \textit{recall} and \textit{precision} metrics for model evaluation.
In Sub-section~\ref{subsec:results}, we present the results of model in both the training and test steps. 
Next, in Sub-section~\ref{subsec:usage_scenario}, we present an usage scenario to attest the applicability of our trained model.
\subsection{Results}
\label{subsec:results}

Table~\ref{table:test_validation} shows the performance of our model in 20-fold cross-validation. 
One can observe that our approach achieved good results in all metrics with a small standard deviation. 
In the test step, our model achieved similar results, as can be seen in Table~\ref{table:test_result}. 

\begin{table}[!ht]
    \centering
    \begin{tabular}{|c|c|c|c|c|}
    \hline
         & \textbf{F1-score} & \textbf{Precision} & \textbf{Recall}   & \textbf{Support} \\ \hline
    \textbf{Appr} & 98.94\%  $\pm$0.14 & 98.40\% $\pm$0.25 & 99.49\% $\pm$0.24 & 1500 \\ \hline
    \textbf{Inap} & 98.93\%  $\pm$0.14 & 99.49\% $\pm$0.24 & 98.38\% $\pm$0.26 & 1500 \\ \hline
    \end{tabular}
    \caption{Evaluation of our approach in 20-fold cross validation}
    \label{table:test_validation}
\end{table}

In a real application of the model, we would choose the model trained with the fold best evaluated. In this case, our best model F1-score for both classes was over 99.00\%.

\begin{table}[!ht]
    \centering
    \begin{tabular}{|c|c|c|c|c|}
    \hline
         & \textbf{F1-score} & \textbf{Precision} & \textbf{Recall}   & \textbf{Support} \\ \hline
    \textbf{Appr}  & 98.95\%  & 98.42\%   & 99.50\%  & 3011 \\ \hline
    \textbf{Inap} & 98.94\%   & 99.49\%   & 98.40\%  & 3011 \\ \hline
    \end{tabular}
    \caption{Evaluation of our approach with test set}
    \label{table:test_result}
\end{table}

Analyzing the F1-score on both Tables \ref{table:test_validation} and \ref{table:test_result}, a remark can be done as for the importance of the F1-score for inappropriate content which has the lowest F1-score of both classes.
It means that less appropriate videos were labeled as inappropriate than inappropriate videos were labeled as appropriate. 
Since this is a censoring-focused model, it is more important to detect inappropriate than to overshoot, mislabeling appropriate videos. Hence, a higher \emph{inappropriate} F1-score is more desirable than a higher \emph{appropriate} F1-score.


\subsection{Usage Scenario}
\label{subsec:usage_scenario}

We propose a usage scenario in which a video file needs to be verified for detecting and censoring any inappropriate scenes on it. 
In such a scenario, one is presented with an hour-long video of an interview. 
To make sure the entire video is safe for uploading on social media, one chooses to use our tool for automatic detecting and censoring inappropriate scenes it may contain.

\begin{figure}[!ht] 
    \centering
    \includegraphics[width=0.6\textwidth]{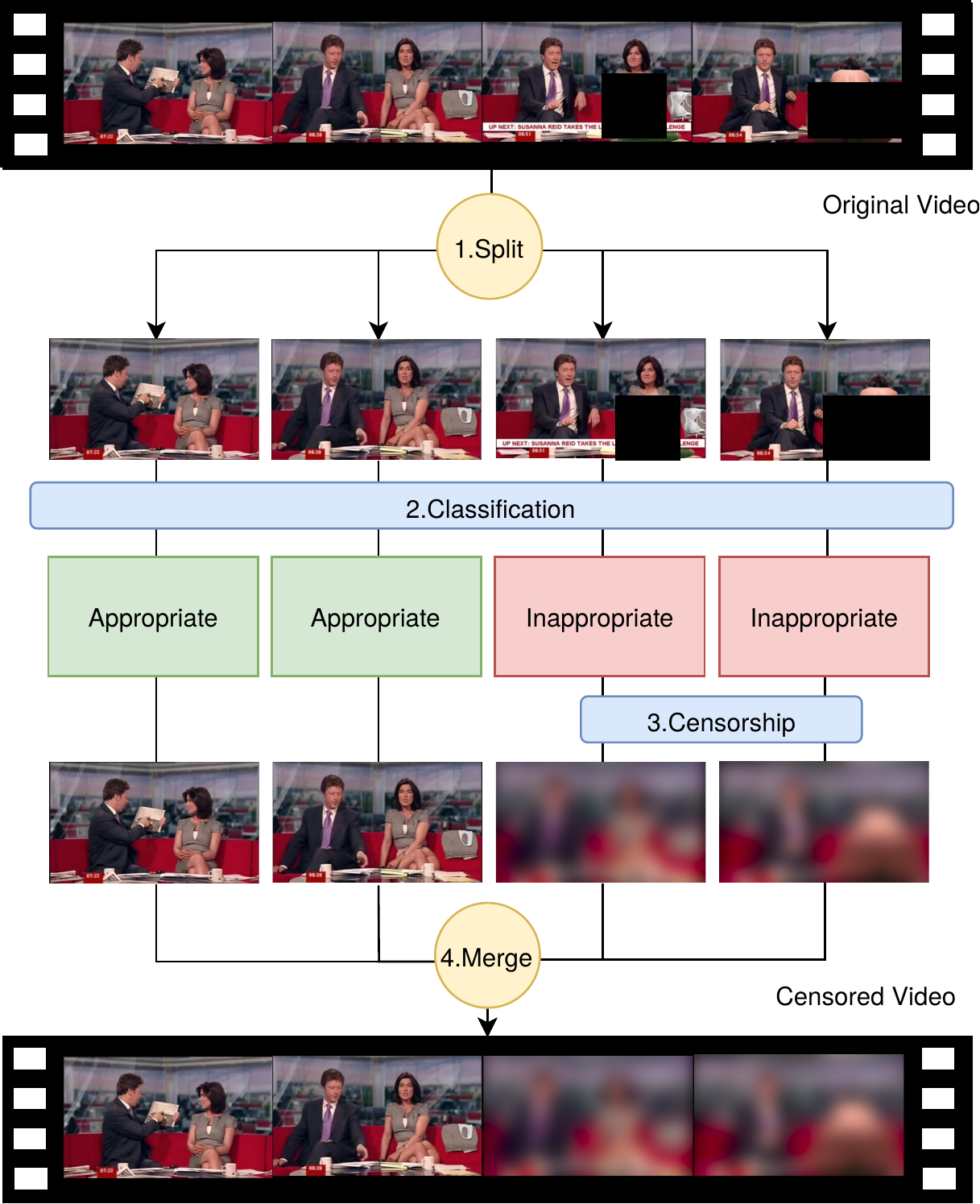} 
    \caption{Censorship Tool use example}
     \label{fig:censorship_example}
\end{figure}

To accomplish this task, our tool first splits the video file into 5 seconds segments.
Then, it uses our classification model to label each video section as either appropriate or inappropriate. Next, it performs the censorship by applying a Gaussian filter on each frame of the segments which are predicted as inappropriate and remove their audio. 
Finally, all resulting segments of the video are merged, generating a new video in which inappropriate scenes are censored. 
Figure \ref{fig:censorship_example} illustrates how the video is split, labeled, censored and merged, resulting in a censored video, accomplishing the task proposed.

\section{Final Remarks}
\label{sec:conclusion}

In this work, we presented a CNN based model for detecting inappropriate scenes in video files.
Our approach achieved high performance with an F1-score of 98.95\% for appropriate videos and 98.94\% for inappropriate videos. 
To attest to the applicability of our proposal, we created a usage scenario of our model, while also creating a tool for automatic censoring inappropriate scenes in video files.

It is important to point out that the tool we proposed only works with video files that are already completely available. 
To extend our tool for working with online content (e.g. video streams and live broadcasts), it would have to be first buffered, generating video segments, processed (what takes some time) and then made available.

\bibliographystyle{unsrt}  
\bibliography{template.bib}

\end{document}